\documentclass[sigconf]{acmart}
\AtBeginDocument{%
  \providecommand\BibTeX{{%
    \normalfont B\kern-0.5em{\scshape i\kern-0.25em b}\kern-0.8em\TeX}}}

\copyrightyear{2022}
\acmYear{2022}
\setcopyright{acmcopyright}
\acmConference[WWW '22 Companion]{Companion Proceedings of the Web Conference 2022}{April 25--29, 2022}{Virtual Event, Lyon, France}
\acmBooktitle{Companion Proceedings of the Web Conference 2022 (WWW '22 Companion), April 25--29, 2022, Virtual Event, Lyon, France}
\acmPrice{15.00}
\acmDOI{10.1145/3487553.3524190}
\acmISBN{978-1-4503-9130-6/22/04}

\begin{document}

\title{Interactions in Information Spread}

\author{Ga\"el Poux-M\'edard\\{\small Under the supervision of Julien Velcin and Sabine Loudcher}}
\email{gael.poux-medard@univ-lyon2.fr}
\affiliation{%
  \institution{Université de Lyon, Lyon 2, ERIC UR 3083}
  \streetaddress{5 avenue Pierre Mendès France}
  \city{Bron}
  \country{France}
  \postcode{69676}
}

\graphicspath{{Source/}}

\renewcommand{\shortauthors}{G. Poux-M\'edard}

\begin{abstract}
Large quantities of data flow on the internet. When a user decides to help the spread of a piece of information (by retweeting, liking, posting content), most research works assumes she does so according to information's content, publication date, the user's position in the network, the platform used, etc. However, there is another aspect that has received little attention in the literature: the information interaction. The idea is that a user's choice is partly conditioned  by the previous pieces of information she has been exposed to. 
In this document, we review the works done on interaction modeling and underline several aspects of interactions that complicate their study. Then, we present an approach seemingly fit to answer those challenges and detail a dedicated interaction model based on it. We show our approach fits the problem better than existing methods, and present leads for future works. Throughout the text, we show that taking interactions into account improves our comprehension of information interaction processes in real-world datasets, and argue that this aspect of information spread is should not be neglected when modeling spreading processes.
\end{abstract}




\maketitle

\begin{figure}
    \centering
    \includegraphics[width=0.40\textwidth]{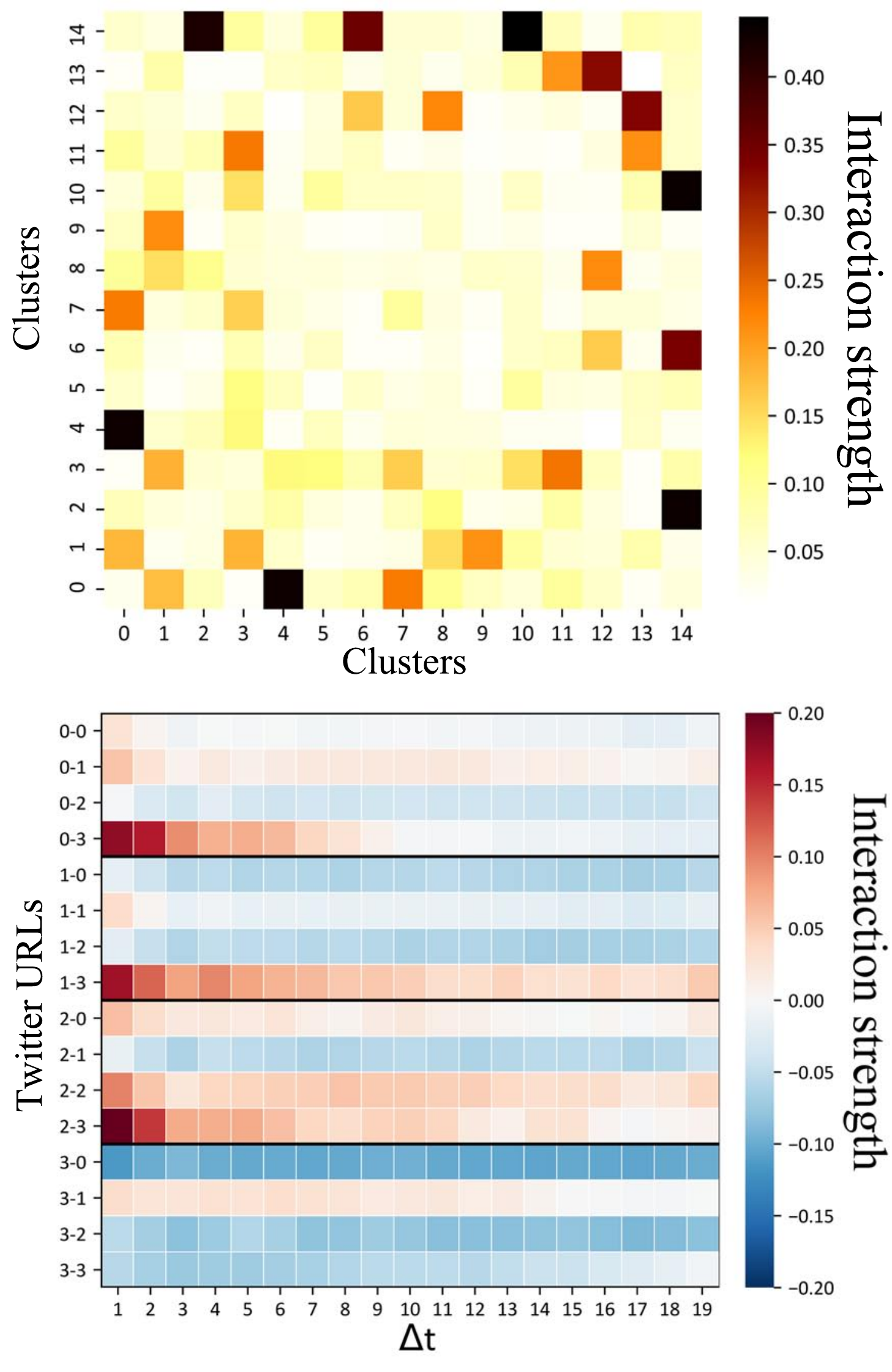}
    \caption{State of the art results on information interaction --- (top) Interactions take place between very specific pairs of clusters; interactions are sparse \cite{Poux2021IMMSBM}. (bottom) Interactions between specific pairs of entities tend to follow an exponential decay in time; interactions do not last \cite{Poux2021InterRate}}
    \label{fig-inter-sota}
\end{figure}
\section{Motivation}
As of today, every minute, approximately 400h of Youtube content, 350.000 tweets and 500.000 Facebook comments are uploaded to the internet. Such amounts of data do not appear randomly. They are generated by users given a context: publication platform, user interests, date of publication, position of the user in a communication network, etc. Many of these aspects have been widely explored in the literature; it has been shown that users interests can be efficiently modeled \cite{Valera2017HDHP,Du2013TopicCascade}, dynamics of publication can be inferred in efficient ways \cite{Du2015DHP}, and methods have been proposed to model the underlying diffusion networks of information diffusion \cite{GomezRodriguez2011NetRate,GomezRodriguez2013InfoPath}. However, an open problem has been independently raised on several occasions, and received little interest over the last decade: how to efficiently model interactions between the spreading \textit{entities}?

Intuitively, we say there is an interaction anytime the combination of entities yields more than their independent sum. Imagine answering a simple question, whose interacting entities are unique words: to the question "Capital France?", the answer is obvious. However, to the bit "Capital?", answers can be very broad; same for the bit "France?". The interaction between those words narrowed the field of possible answers; they are interacting. This idea can easily be generalized to retweets (does a user retweet A only because of A, or also because of tweet B she has seen before?), online buying (does the user buy A because she likes it, or because she also bought B before?), and in general to online publication of documents (what are the bits of information making a user publish this post in particular on this particular platform?).

The understanding of interactions in information spread should provide a more exhaustive description of the publication process on the internet, as well as provide leads on how users react in a given context. At a time where three quarters of the world population has an online identity\footnote{\url{https://en.wikipedia.org/wiki/Global_Internet_usage}}, the understanding of these mechanisms could find applications anytime an internet user has to make a choice: click, buy, like, comment, etc. One could think of nudging users towards healthier internet behaviours on forum or better consumption habits on online retail websites for instance.

The goal of this document is to briefly review the challenges that arise when it comes to modeling interactions in information spread --interactions are sparse and short. We then present a candidate methodology to deal with them, based on both Bayesian clustering and temporal point-processes. We show this method has flaws and propose a way to correct them. Our method allows to get better results on several datasets. Finally, we present leads to extend the existing framework and discuss research perspectives on interactions modeling.

\section{Problem}
\subsection{Interactions are sparse}
In a first attempt to explicitly model interactions in a real-world dataset, the authors in \cite{Myers2012CoC} proposed to apply a classical block model to Twitter data. The goal is to predict the probability that a link will be retweeted, given a history of links that appeared before it. The method makes a central assumption: the interaction between tweets is a small modulation of a base retweet probability, which is precomputed as the total frequency of this tweet being retweeted. On this basis, the authors show that interaction modeling indeed helps to improve the predictive performances of the model. The analysis of the inferred interaction terms has led them to the following conclusions: most interactions are small, and interactions are slightly negative overall (meaning the lower the base probability of a retweet). By running their model for different sizes of prior exposures (or history), they also conclude that users rely mostly on a short-term memory, which correlates previous findings on users modeling \cite{Guttieres2016MrBanks,Poux2021MMSBMMrBanks}.

However, more recent works proved the core-hypothesis of this modeling false \cite{Poux2021IMMSBM}. By precomputing the base probability of a retweet as the total number of times a tweet has been retweeted divided by the number of times it appeared, the authors in \cite{Myers2012CoC} already account for interactions. Imagine that interactions lead to a shift of $\Delta p$ on the true base probability of a retweet $p$, and that this interaction happens in a fraction $f$ of all observations of a given tweet being retweeted. The base probability as defined in \cite{Myers2012CoC} then equals $p(1-f) + (p+\Delta p) f = p + f \Delta p$, which is likely not to equal the actual probability of a retweet in the absence of interaction $p$. Therefore, defining interaction according to $p$ is wrong. The base probability for a retweet needs to be inferred by the model at the same time as the contribution of interactions to be properly defined.

The authors in \cite{Poux2021IMMSBM} then propose a mixed membership stochastic block model (MMSBM) in order to jointly infer the base probability of an entity spreading, as well as the interaction terms that modulate it. The idea is to group entities into clusters according to the way those clusters interact together. The authors in \cite{Poux2021IMMSBM} study the importance of interactions in 4 different datasets. The conclusions are as follow: \textbf{modeling interactions improve predictive results, most interactions are small, and significant interactions are sparse}, meaning that they take place only between rare specific pairs of clusters. This result for the Twitter dataset is reported on Fig.~\ref{fig-inter-sota}-top.

\subsection{Interactions are short}
An important aspect of interactions that \cite{Poux2021IMMSBM} does not consider is time. The position of entities in the considered history is not considered. In \cite{Myers2012CoC}, the order of appearance is taken into account, but each temporal slice is modeled as independent from the others. 

In \cite{Poux2021InterRate}, the authors propose to study the evolution of the interaction term in continuous time, using a convex model based on survival theory. This framework allows to infer both a time-independent probability of a retweet \textit{and} a linear combination of time-dependent kernel functions that account for temporal evolution of the interaction terms. The data considered is similar as in \cite{Myers2012CoC,Poux2021IMMSBM} as it consists of pairs of entities (typically tweets) associated to a given action (typically retweet), separated by a time $\Delta t$ in a user's history. Note that this model does not have any clustering component; the interaction is studied only between the most interacting pairs of entities for every of the 3 considered datasets.

The results of \cite{Poux2021InterRate} are as follow: \textbf{interactions magnitudes follow an exponential decay in time, and therefore do not last in time}. Besides, the model recovers previous findings on interactions: interactions are sparse, and their maximum magnitude is small. This result is reported Fig.~\ref{fig-inter-sota}-bottom.

\section{State of the art}
\subsection{Historical approach to dynamic clustering}
We reviewed the main challenges associated with interactions modeling. First, interactions take place between specific pairs of clusters, and even more specific pairs of entities; to be able to spot them using non-convex models requires a \textbf{clustering aspect}. Second, interactions quickly fade in time, meaning that the information clusters must also take into account a \textbf{temporal aspect}.

The use of temporal dimension in documents clustering has been studied several occasions; a notable spike of interest happened in 2006. Many authors tackled the problem of inferring time-dependent clusters using models based on LDA \cite{Blei2006DynamicTopicModel,Wang2006TopicsOverTime}. The idea is to sample the data provided to each model with a temporal function; the data can be selected using a temporal sliding window, or a temporal data sampling function. Time is not explicitly modeled. Most of these models are parametric, meaning the number of clusters is fixed at the beginning of the algorithm. In 2008, A. Ahmed \textit{\& al} proposed the Recurrent Chinese Restaurant Process (RCRP) as an answer to this problem \cite{Amr2008RCRP}. Instead of considering a fixed-size dataset, this model can handle a stream of documents arriving in chronological order, and the number of clusters is automatically updated. In this model, time is still not modeled explicitly (data is split into episodes to capture the temporal aspect of cluster formation). A later version of the model from 2010, the Distance-Dependent Chinese Restaurant Process (DD-CRP), tries to alleviate this approximation by replacing fixed-time episodes with a continuous-time sampling function \cite{Blei2010DDCRP}. However, the model is still not designed to explicitly model temporal information in a continuous-time setting. The slicing of the considered data introduces a consequent bias in the modeling.

\subsection{Dirichlet-Hawkes process (DHP)}
In 2015, N. Du \textit{\& al} \cite{Du2015DHP} answered this problem by developing the Dirichlet-Hawkes Bayesian prior, that can be used to model the appearance of events in a continuous-time setting. The key idea is to replace the counts in a Dirichlet process prior by the intensity function of a Hawkes process. 

Explicitly, given a standard Bayesian textual clustering model: 
\begin{equation}
    \label{eq-simpleBayes}
    \underbrace{P(\text{cluster} \vert \text{text, time})}_{\text{Posterior probability}}
        \propto \underbrace{P(\text{text} \vert \text{cluster})}_{\text{Textual likelihood}} 
        \times \underbrace{P(\text{cluster} \vert \text{time})}_{\text{Temporal prior}}
\end{equation}
the authors propose the following form for the temporal prior:
\begin{equation}
\label{eq-DHP}
    P(C_i = c\vert t_i, \lambda_0, \mathcal{H}_{<t_i,c}) = 
    \begin{cases}
    \frac{\lambda_c(t_i)}{\lambda_0 + \sum_{c'} \lambda_{c'}(t_i)} \text{ if c$\leq$C}\\
    \frac{\lambda_0}{\lambda_0 + \sum_{c'} \lambda_{c'}(t_i)} \text{ if c=C+1}
    \end{cases}
\end{equation}
where $C_i$ is the cluster allocation among $C$ clusters of the $i^{th}$ document appearing at time $t_i$, $c$ a random variable, $\mathcal{H}_{<t_i,c}$ the date of all documents within cluster $c$ published before $t_i$, $\lambda_c(t)$ the intensity of cluster $c$ at time $t$, and $\lambda_0$ a parameter. 

\begin{figure}
    \centering
    \includegraphics[width=0.45\textwidth]{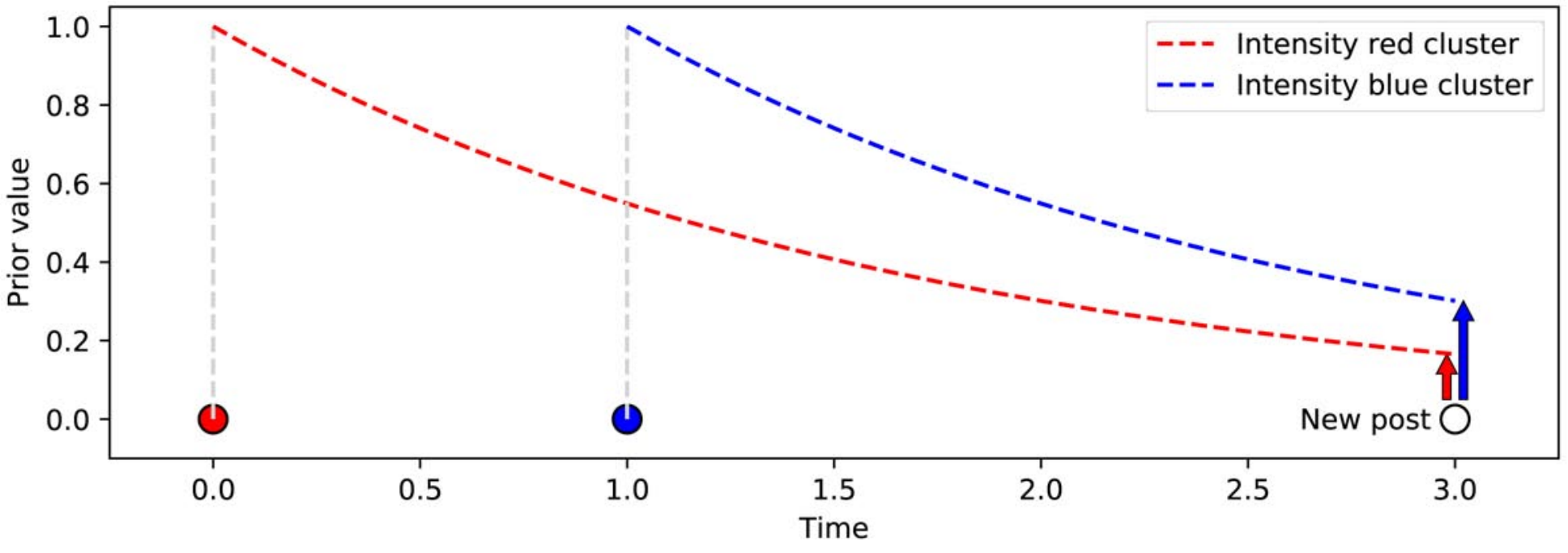}
    \caption{Illustration of the Dirichlet-Hawkes prior --- Each cluster is associated to a temporal intensity function. The prior probability (Eq.\ref{eq-DHP}) of a new observation belonging to a cluster is linearly proportional to this cluster's intensity.}
    \label{fig-illust-dhp}
\end{figure}

The functions $\lambda_c(t)$ need to be inferred as the intensities of a Hawkes process \cite{Rizoiu2017Hawkes}, and represent the instantaneous probability that a document appears at time $t$. Note that in Eq.\ref{eq-DHP}, the prior probability that a new document belongs to a given cluster depends on the intensity of this cluster at a time $t$: \textbf{clusters are self-stimulated}. In the problematic of information interaction, it means that pieces of information can only interact with other pieces of information from the same cluster. An example of a situation where a new observation gets allocated to any of the existing clusters is illustrated Fig.~\ref{fig-illust-dhp}.

The resulting Dirichlet-Hawkes process (DHP) is then used as a prior for clustering documents appearing in a continuous-time stream. The inference is realized with a Sequential Monte-Carlo (SCM) algorithm. Following DHP, two articles have been published extending the idea: the Hierarchical Dirichlet Hawkes process (HDHP) \cite{Valera2017HDHP} in 2016 and Indian Buffet Hawkes process in 2018 \cite{Tan2018IBHP}. Recently, \cite{Kapoor2018BayesianNonparametrics} underlined that this formulation suffers from a vanishing prior problem ($\lambda_c(t)$ can go to 0), and proposes a small procedure in order to avoid it, without modifying the core idea of the prior.

\section{Proposed approach}
\subsection{Improvements over DHP}
A common feature of all the models we mentioned built on DHP is that they use a non-parametric Dirichlet process (DP) prior or variations built on it, such as DHP and HDHP. Yet, on several occasions, it has been pointed out that there are no specific reasons to use this process in particular and that alternative forms might work better depending on the dataset \cite{Welling2006AlterDP,Wallach2010UnifP}. In \cite{Welling2006AlterDP}, the author relaxes several properties associated with DP and shows that alternative priors are an equally valid choice in Bayesian modeling. In \cite{Wallach2010UnifP}, the authors derive the Uniform process (UP) and show that it performs better on a document clustering task. In \cite{Poux2021PDP}, the authors generalize UP and DP within a more general framework, the Powered Dirichlet process (PDP), and show it performs better than DP on several datasets.

Moreover, it has recently been highlighted that DHP does not work well when the textual information within documents conveys little information, that is when the text is short \cite{Yin2018ShortTextDHP} or when vocabularies overlap significantly. To answer this problem, the authors develop an approach based on Dirichlet process mixtures, which is not designed for continuous-time document streams -- the temporal aspect comes from a sampling function as in \cite{Amr2008RCRP,Blei2010DDCRP}. 
There are other limiting cases for DHP, for instance when temporal information is conveys little information (few observations, entangled dynamics) or when documents within textual clusters do not follow a unique temporal dynamic. To overcome those limitations, we develop the Powered Dirichlet-Hawkes process in the next section.

\subsection{The model}
Taking back Eq.\ref{eq-DHP}, we note that the linear dependence of the prior on $\lambda_c(t)$ is justified only by analogy with the standard Dirichlet process. At this stage, we believe that a small modification of Eq.\ref{eq-DHP} can lead to great variations in the comprehension of a dataset. In particular, making the prior more or less dependent on the temporal dimension (in the same way that \cite{Wallach2010UnifP,Poux2021PDHP} makes the DP more or less dependent on the ``rich-get-richer'' hypothesis) could lead to clusters that are more text or time orientated. By replacing the standard Dirichlet process in \cite{Du2015DHP} by the Powered Dirichlet process from \cite{Poux2021PDP}, we derive the Powered Dirichlet-Hawkes process\cite{Poux2021PDHP}:
\begin{equation}
\label{eq-PDHP}
    P(C_i = c\vert r, t_i, \lambda_0, \mathcal{H}_{<t_i,c}) = 
    \begin{cases}
    \frac{\lambda_c^r(t_i)}{\lambda_0 + \sum_{c'} \lambda_{c'}^r(t_i)} \text{ if c$\leq$C}\\
    \frac{\lambda_0}{\lambda_0 + \sum_{c'} \lambda_{c'}^r(t_i)} \text{ if c=C+1}
    \end{cases}
\end{equation}

\begin{figure}
    \centering
    \includegraphics[width=0.40\textwidth]{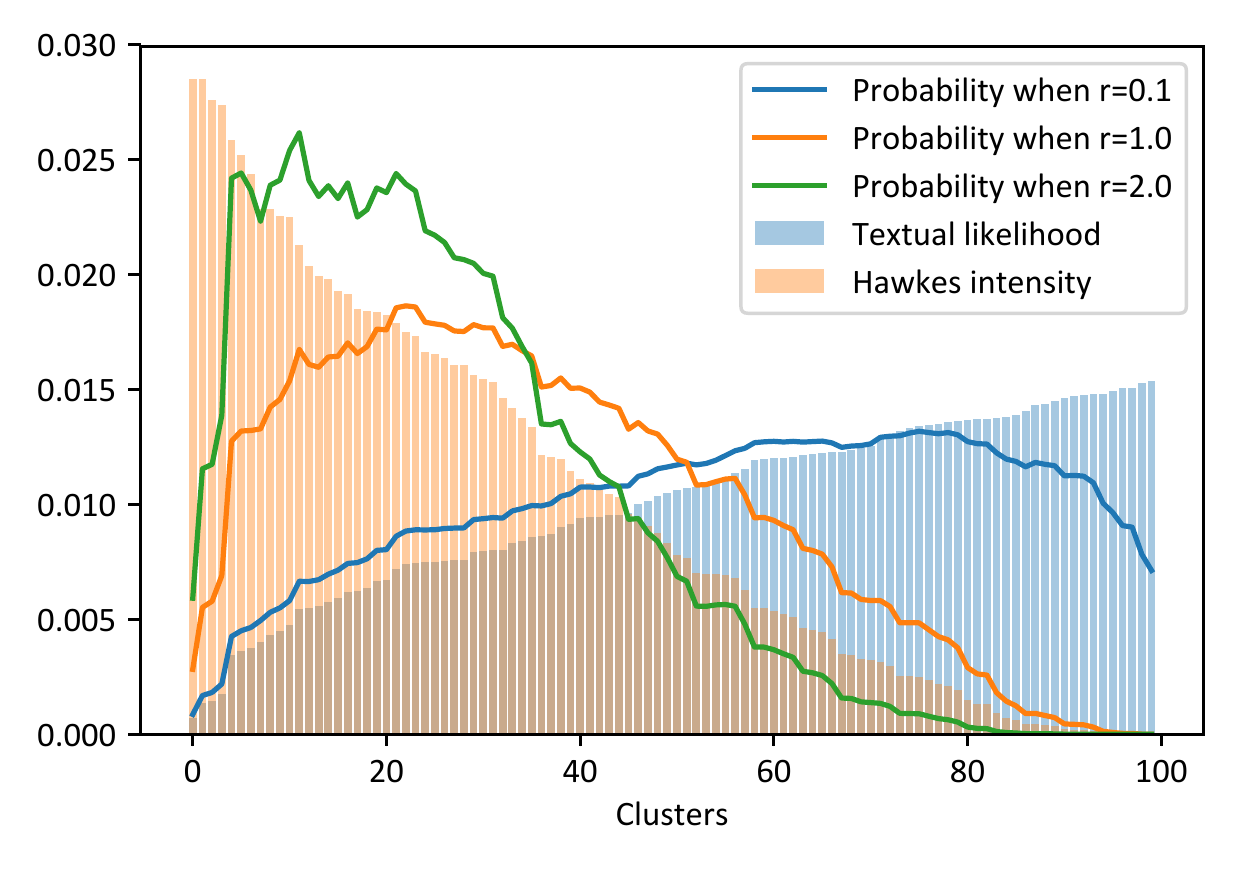}
    \caption{Effect of the parameter $r$ in PDHP on the posterior probability (Eq.\ref{eq-simpleBayes} using the prior in Eq.\ref{eq-PDHP}) for a new observation to belong to each cluster.}
    \label{fig-illust-effect-dhp}
\end{figure}

We show in Fig.~\ref{fig-illust-effect-dhp} what this change implies for a standard Bayesian textual clustering model. The value of $r$ tunes how much we rely on either the textual or the temporal information every time the algorithm chooses a cluster for a new document.

The algorithm used for inference is a sequential Monte-Carlo; it is the same as described in \cite{Du2015DHP,Valera2017HDHP,Poux2021PDHP}. It is especially fit for modeling data streams, as it considers data sequentially. The underlying idea is that for each new observation, a number of \textit{particles} $N_{part}$ will make a cluster allocation hypothesis according to Eq.\ref{eq-simpleBayes}, update the cluster's intensity function $\lambda_c(t)$, and pass on to the next document. When the likelihood of a particle $L_p$ goes under a given threshold $\omega$, the particle is discarded and replaced by a more likely one. In our case, we consider 8 particles and a threshold of $\omega=\frac{\sum_p L_p}{2N_{part}}$

\section{Methodology and results}
\subsection{Data generation}
We simulate cases where only two clusters are considered. Each cluster has its own vocabulary distribution over 1~000 words and its associated intensity function $\lambda_c(t)$.
We first simulate one independent Hawkes process per cluster using the Tick Python library \cite{Bacry2017Tick}. The processes are stopped at time $t=1500$, which makes a rough average of 7~000 events per run. Then we associate each simulated observation with a sample of 20 words drawn from the corresponding cluster's word distribution. 

We evaluate our results according to the Normalized Mutual Information score (NMI), which is a standard metric to compare the results of different clustering algorithms when the number of clusters can vary.

\begin{figure}
    \centering
    \includegraphics[width=0.40\textwidth]{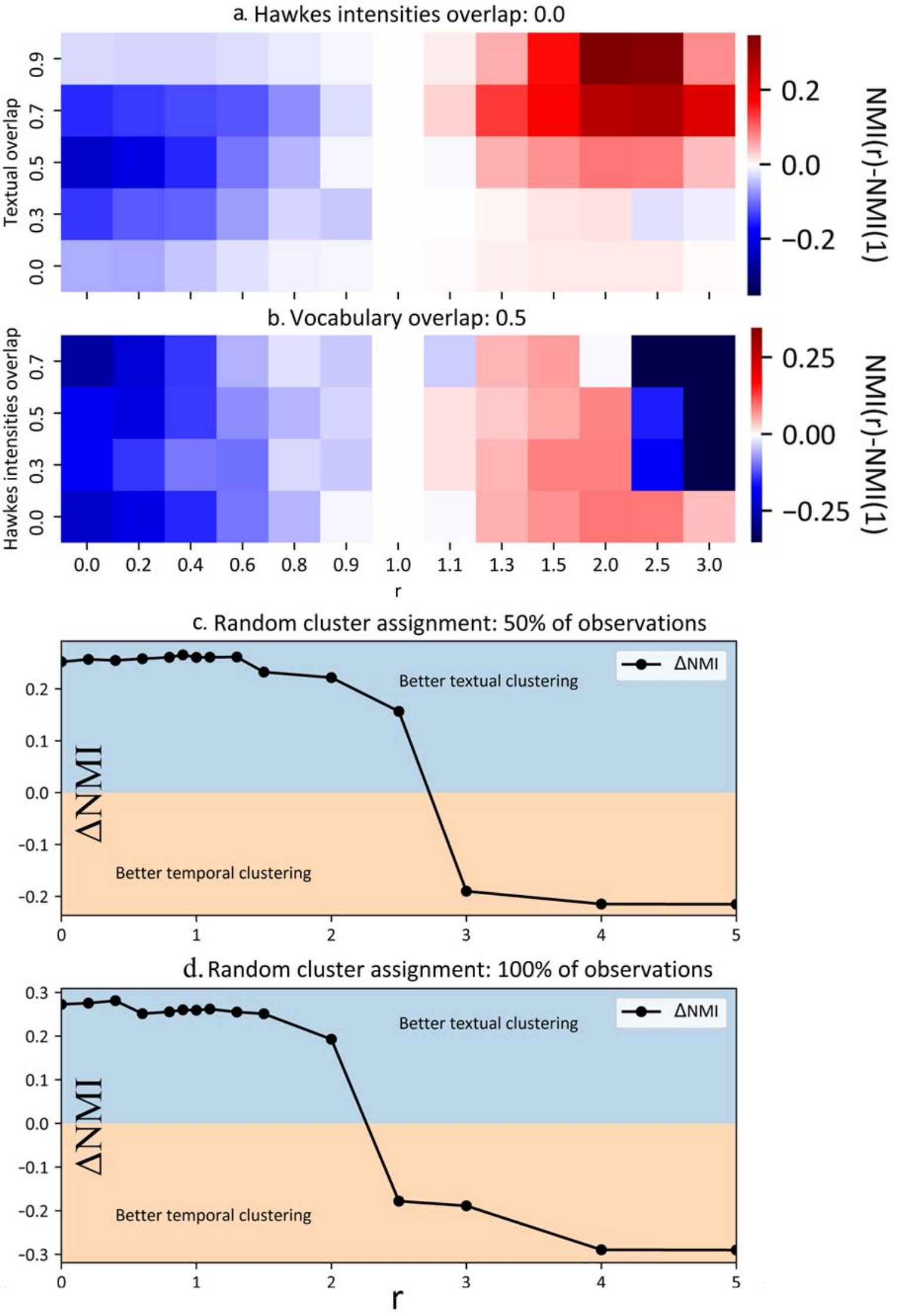}
    \caption{Results of PDHP when textual and temporal information are not fully reliable (a,b) and when textual content is decorrelated from temporal dynamics (c,d).}
    \label{fig-res-synth}
\end{figure}

\subsection{Intra-cluster interactions...}
\subsubsection{...with unreliable information}
A first lead to explore how information interact in time is to consider the case where clusters are self-stimulated. However, \cite{Yin2018ShortTextDHP} showed that the approach in \cite{Du2015DHP} is not fit for cases where textual information is weak or unreliable (e.g. Twitter) or when temporal dynamics are highly entangled or when textual information is sufficient (e.g. long documents or real-world process at large scales). In order to confirm that our modeling in \cite{Poux2021PDHP} answers those limitations, we run experiments for various values of informativeness of either textual or temporal content. 

To do so, we will make the clusters' vocabulary distributions and temporal intensity functions overlap. Given two distributions $\mathcal{A}(x)$ and $\mathcal{B}(x)$, the overlap is defined as $\frac{\int_x min(\mathcal{A}(x),\mathcal{B}(x)) dx}{\int_x \mathcal{A}(x)+\mathcal{B}(x) dx}$.

In the case of vocabulary overlap, we shift the words distribution so that they have a certain fraction of shared words. In the case of temporal overlap, we first simulate each cluster's Hawkes process, and compute the intensity function of the whole resulting dataset. Then, we shift each process realization on the temporal axis so that the overlap reaches the desired value. We evaluate the models according to the documents clusters allocations compared to the true clusters from the data generation process. We report the average improvement for every of the 20 datasets considered in each overlaps configuration, for various values of $r$.

Our results are shown in Figs.~\ref{fig-res-synth}a-b. PDHP allows better performances comparing to DHP (r=1) and UP (r=0) when textual information is unreliable (high textual overlap), up to a gain of +0.3 NMI w.r.t. DHP (Fig.~\ref{fig-res-synth}a). In more realistic situations (average textual and temporal overlaps), PDHP allows better results (up to +0.2 NMI, Fig.~\ref{fig-res-synth}b). Our method is therefore able to accurately uncover which cluster triggered a new document, that is to which interaction chain this document belongs, in the case of weak temporal and textual information.

\subsubsection{... with decorrelated content and dynamics}
Another limiting situation of \cite{Du2015DHP} is when textual documents about a same topic do not follow similar dynamics. Picture a famous and an infamous newspapers publishing the exact same article; these articles would not follow the same dynamics despite an identical textual content.

In order to explore the case where identical textual contents do not follow identical temporal dynamics, we first generate the whole dataset with null temporal and textual overlaps. Then, we randomly select a given fraction of the generated observations, and resample their associated vocabulary according to a randomly selected cluster. 
We have two different possible clustering: textual (how well the model recovers the vocabulary a document's content has been drawn) and temporal (which intensity function made this document appear where it is). In this case, we report the difference between the textual clustering NMI and the temporal clustering NMI: $\Delta NMI = NMI_{text}-NMI_{time}$.

Our results are shown Figs.~\ref{fig-res-synth}c-d. When textual content is decorrelated from publication dynamics (which is more in line with what happens in real-world processes), tuning $r$ allows to consistently recover one clustering or the other. That means we can successfully recover clusters based on either textual similarity or on the temporal interaction between documents, or a mixture of both.

\subsection{Ongoing: Inter-cluster interactions}
As for now, we presented introductory results about self-interacting clusters. However, in order to get the full picture, we cannot only consider the case where only one given cluster interacts with incoming documents. Instead, we need to consider the interaction between all existing clusters. This embeds naturally in the proposed approach \cite{Du2015DHP,Poux2021PDHP} by considering a multivariate Hawkes process instead of a uni-variate one \cite{Rizoiu2017Hawkes}. In implies redefining the numerator in Eq.\ref{eq-PDHP} from $\lambda_c(t)$ to $\sum_{c'} \lambda_{c'\rightarrow c}(t)$, where $\lambda_{c'\rightarrow c}(t)$ is the temporal influence of cluster $c'$ on cluster $c$ at a time $t$. The challenge here is to preserve the efficiency of the proposed algorithm while the model's complexity increases in a polynomial fashion with the number of inferred clusters.

\subsection{Future: User-level inter-cluster interactions}
As for future works, we believe it is possible to go further by combining temporal point processes to Dirichlet processes. While \cite{Du2015DHP} opened the door to this union, it only explored a small part or possible applications. In particular, we think of the works of Gomez-Rodriguez on underlying network inference \cite{GomezRodriguez2011NetRate,GomezRodriguez2013InfoPath} and extensions \cite{Du2013TopicCascade,Wang2012Monet}. In \cite{GomezRodriguez2013SurvivalAnalysis}, he showed that all these models can be formulated in terms of a counting temporal point process, which naturally allows to substitute any such modeling to the Hawkes process in Eq.\ref{eq-DHP} and Eq.\ref{eq-PDHP}. In this case, we would be able to study the interaction between pieces of information at the user-level, providing a deeper understanding of information interaction in real-world processes.

\section{Conclusions}
In this document, we first motivated the need to model the interaction between pieces of information in order to get a better understanding of real-world spreading processes. We then highlighted the main challenges that arise in this perspective using recent published works. We proposed to address the problem by considering a recent approach that seems to fit to answer the challenges of interaction modeling, despite having its own flaws. We proposed a simple method to overcome these, and conducted systematic experiments to demonstrate how our methodology works at doing so. Finally, we proposed leads for further improving our understanding of real-world information interaction modeling.


\bibliographystyle{ACM-Reference-Format}
\bibliography{main}

\end{document}